\documentclass[12pt]{article}
\usepackage{fullpage}
\usepackage{amsmath}
\usepackage{amssymb}
\usepackage{epsfig}
\usepackage{feynmf}
\usepackage{subfigure}

\begin{document}

\begin{center}
{\Large \bf Probing new physics in rare charm processes\footnote{Talk at the Cairo International Conference on 
High Energy Physics, 9-14 January 2001, Cairo, Egipt}}\\
\vspace{1cm}
S. Prelovsek

\vspace{.3cm}

{\it Department of Physics, University of Ljubljana and J. Stefan Institute, Ljubljana, Slovenia
\\E-mail: sasa.prelovsek@ijs.si}

\end{center}

\vspace{1cm}

\centerline{\bf ABSTRACT}

\vspace{0.3cm}

We explore the sensitivity of the flavour changing neutral currents among $c$ 
and $u$ quarks in the context of the Minimal Supersymmetric Standard Model. The standard model rate for $c\to u\gamma$ can be enhanced  by  up to two orders of magnitudes in general MSSM, while the $c\to u l^+l^-$ rate can be enhanced by at most  a factor of three. 
The most suitable hadronic observable to probe the $c\to u\gamma$ transition is the decay $B_c\to B_u^*\gamma$   or the difference in decay rates of $D^0\to \rho^0\gamma$ and $D^0\to \omega\gamma$, since they are relatively free of the long distance contributions. The $c\to ul^+l^-$ transition may be probed in $D\to \pi l^+l^-$ decays at high high di-lepton mass  and we predict  the long distance contributions for these modes.  

\vspace{2cm}

\section{Introduction}

Flavour changing neutral (FCN) currents  occur in the standard model only 
at the loop level where they are suppressed by the GIM mechanism.  The FCN transitions among the up-like quarks $u$, $c$ and $t$  are especially 
rare  due to the small masses of the intermediate 
down-like quarks and only the upper experimental limits for this processes are 
available at present \cite{PDG,radiative.exp,vll.exp,pll.exp}. In the present paper we turn to the FCN transitions among the up-like quarks with the highest  branching ratio in the standard model, namely the $c\to u\gamma$ and $c\to ul^+l^-$ transitions. We do not discuss the decays $c\to ug$ and $c\to u\nu\bar \nu$ which would hardly lead to a distinctive experimental signature, or the helicity suppressed decay $D^0\to l^+l^-$. The $D^0-\bar D^0$ mixing has been thoroughly studied with the predictions compiled in \cite{nelson} and the new exciting experimental data presented in \cite{mixing}.

The FCN processes are suitable as probes for
the popular low energy supersymmetry,  
which has various new sources of flavour 
violation. Such investigation may reveal the flavour 
structure of the soft supersymmetry breaking terms and shed light on the 
mechanism of SUSY breaking. In this note we explore the sensitivity of the $c\to u\gamma$ \cite{sasa} and $c\to ul^+l^-$ \cite{pll} decay rates 
in the framework of the  minimal supersymmetric standard model (MSSM). 

The effect of new physics on the quark decay of interest may be overshadowed by the more mundane effects of the 
long distance (LD) dynamics in hadron decays.  The least contaminated LD hadronic observable to probe $c\to u\gamma$ 
transition
is found to be the decay $B_c\to B_u^*\gamma$ \cite{bc} or the difference of the decay rates $D^0\to \rho^0\gamma$ and $D^0\to \omega \gamma$ \cite{FPSW}.  Probing $c\to ul^+l^-$ might be possible 
at high $m_{ll}$ in $D\to \pi l^+l^-$ decay \cite{pll}. 

\vspace{0.1cm}

The SM and MSSM predictions for the $c\to u\gamma$ and $c\to ul^+l^-$ rates are given in Section 2, while the interesting hadronic observables are discussed in Section 3. Conclusions are drawn in Section 4.  

\section{${c\to u\gamma}$ and $c\to ul^+l^-$ decays}

\subsection{Standard model}

The $c\to u\gamma$ amplitude  is strongly GIM suppressed at one loop electroweak order, ${\cal A}(c\to u\gamma)\propto \sum_{d,s,b}V_{cq}^*V_{uq}m_q^2$, leading to a branching ratio of only $\sim 10^{-18}$. The two-loop QCD corrections to the responsible Willson coefficient $c_7$ were found to be large \cite{GHMW}, leading to
\begin{equation}
\label{cugamma.br.sm}
Br^{SM}(c\to u\gamma)\simeq 3\times 10^{-8}~.
\end{equation}

\begin{figure}[!htb]
\begin{center}
\includegraphics[scale=0.5]{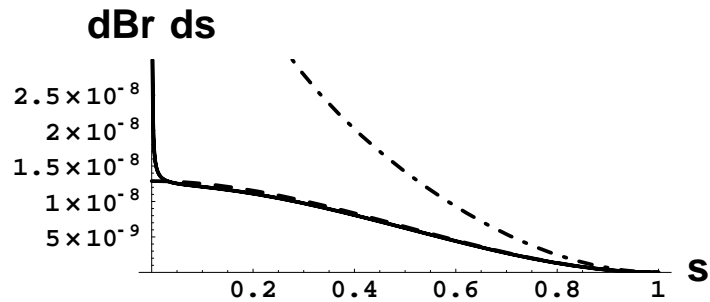}
 \caption{The differential branching ratio $dBr(c\to ul^+l^-)/ds$ 
with $s=m_{ll}^2/m_c^2$: the dashed line denotes the one-loop standard 
model prediction, while the solid line incorporates also the QCD 
corrections to the Willson coefficient $c_7$ \cite{GHMW}.  The  
best possible enhancement of the $c\to ul^+l^-$ rate in the general MSSM  is displayed by the
dot-dashed line, where the mass insertions are taken  at their maximal values
(\ref{insertionlr}), (\ref{insertionll}). Masses of squarks and gluinos are taken at   $250$ GeV and  
$\alpha_s=0.12$.  
\label{fig.sd}}
\end{center}
\end{figure}

The dominant contribution to  $c\to ul^+l^-$   at  the one-loop electroweak level is {\it not} strongly GIM suppressed \cite{pll,vll}, ${\cal A}(c\to ul^+l^-)\propto V_{cs}^*V_{us}\ln[m_s^2/m_d^2]$, leading to
\begin{equation}
\label{cull.br.sm}
Br^{SM}(c\to ul^+l^-)\simeq 6\times 10^{-9}~.
\end{equation}
The two-loop QCD corrections to $c_7$ \cite{GHMW} do not  affect the rate significantly, as shown in Fig. \ref{fig.sd}.
 
\subsection{Minimal supersymmetric standard model}

New sources of flavour violation are present in the minimal 
supersymmetric
standard
model (MSSM) and these  depend crucially on the mechanism of 
the
supersymmetry
breaking. The schemes with flavour-universal soft-breaking terms lead to
contributions proportional to $\sum_{q=d,s,b}V_{cs}^*V_{us}m_q^2$ \cite{duncan.wyler} and have
negligible effect on the $c\to u\gamma$ and $c\to ul^+l^-$ rates \cite{sasa,masiero.cugamma}. Our
purpose
here is to explore the largest possible enhancement of the $c\to u\gamma$ and $c\to ul^+l^-$
rates
in MSSM   
with unbroken R-parity and arbitrary non-universal soft 
breaking terms.

\begin{figure}[!htb]
\begin{center}
\input{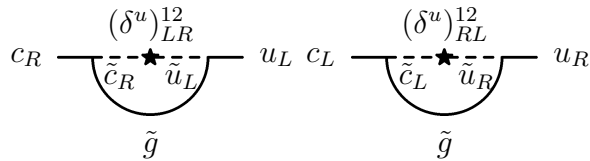}
 \caption
{ The diagram giving the dominant supersymmetric contribution to $c\to u\gamma$ and $c\to u\gamma\to ul^+l^-$ decays in general MSSM, given the constraints (\ref{insertionlr}) and (\ref{insertionll}) on the mass insertions.\label{fig.insertionlr}}
\end{center}
\end{figure}

The first analysis of the $ {c\to u\gamma}$  decay in MSSM has been presented in \cite{masiero.cugamma}, where only the additional penguin diagrams with gluinos in the loop were considered. The complete analysis, including the QCD corrections, for the penguin diagrams with gluinos, charginos and neutralinos in the loop was presented in \cite{sasa}. We choose the super-CKM basis for squarks, in which the quark - squark - gaugino vertex has the same flavour structure as the quark - quark - gauge boson vertex, and undertake the mass insertion approximation.  In this framework, the flavour change is provided by the non-diagonality of the squark propagators and is parameterized by $(\delta^{q}_{ij})_{AB}=({\cal M}^2_{ij})^q_{AB}/m_{\tilde q}^2$ with $A,B=L,R$ and squark mass matrix  squared $({\cal M}^2)_{AB}^q$. The  
 diagram, that eventually turns out to give the dominant contribution,  is shown in Fig. \ref{fig.insertionlr}.

The mass insertions are free parameters in a general MSSM. The strongest
upper
bound on  $(\delta_{12}^u)_{LR}$ comes by requiring that the minima of the scalar potential do not break charge or color, and that they are bounded
from
bellow \cite{sasa,CD}
\begin{equation}
\label{insertionlr}
 |\delta_{12}^u|_{LR}~,~|\delta_{12}^u|_{RL}\leq 0.0046~\qquad{\rm for}\quad
M_{sq}=250\ {\rm GeV}.
\end{equation}
The insertions $(\delta_{12}^u)_{LL}$ and $(\delta_{12}^u)_{RR}$ 
can be
bounded
by saturating the experimental upper bound $\Delta m_D<4.5\times 10^{-14}$
GeV
\cite{mixing} by the gluino exchange \cite{sasa,gabbiani}. Since we are interested in exhibiting the largest
possible  enhancement, we saturate $\Delta m_D$ by $(\delta_{12}^u)_{LL}$
\cite{sasa,gabbiani}
\begin{equation}
\label{insertionll}
 |\delta_{12}^u|_{LL}~\leq 0.03\qquad{\rm for}\quad
M_{sq}=M_{gl}=250\ {\rm GeV}
\end{equation}
and set $(\delta_{12}^u)_{RR}=0$.  

\begin{figure}[!htb]
\begin{center}
\includegraphics[scale=0.6]{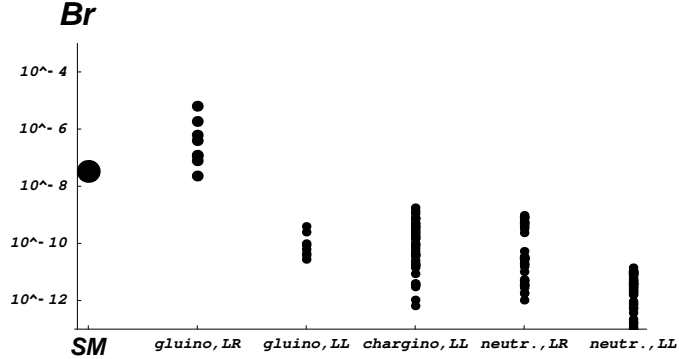} 
 \caption{The scatter plot of predicted $c\to u\gamma$ branching ratios for various contributions in MSSM.  
The QCD-corrected \cite{GHMW} standard model rate (\ref{cugamma.br.sm})  is represented by the  dot on the left. The remaining dots represent the rates arising solely from gluino, chargino and neutralino exchange; the contributions arising from $LL$ and $LR$ insertions are shown separately. The mass insertions are taken at their maximal values (\ref{insertionlr}) and (\ref{insertionll}), while other parameters are taken in the ranges 
$250$ GeV$\leq m_{\tilde q},m
_{\tilde g}\leq 1000$ GeV,  $\alpha_s\!= 0.12$, 
$2.5\leq tg\beta\leq 30$,  $100~{\rm GeV}\leq |\mu|\leq 300$ GeV and 
$m_{\chi_1^+}>90$ GeV. \label{fig.cugamma}}  
\end{center}
\end{figure}

The scatter plot of the $c\to u\gamma$ branching ratio that  result from the 
gluino, chargino and neutralino  contributions  is shown in Fig. \ref{fig.cugamma}, where the mass insertions are taken at the maximal values of (\ref{insertionlr}) and (\ref{insertionll}).
If $(\delta^u_{12})_{LR}$ is close to the upper value in (\ref{insertionlr}), the  gluino exchange diagram in Fig. \ref{fig.insertionlr}   can enhance the standard model 
rate (\ref{cugamma.br.sm}) by up to two orders of magnitudes    
\begin{equation}
Br(c\to u\gamma)_{LR}\simeq 6\times 10^{-6}~.
\end{equation}
In contrast, the the contributions from chargino and neutralino remain at least an order of magnitude bellow the QCD corrected standard model rate.

\vspace{0.2cm}

Based on experience from $c\to u\gamma$, we consider only the gluino exchange diagrams with a single mass insertion for the case of ${c\to ul^+l^-}$ decay. The dominant contribution again arises from the diagrams in Fig. \ref{fig.cugamma} with the lepton pair attached to the photon. The best possible enhancement in general MSSM is shown by dot-dashed line in Fig. \ref{fig.sd}. The effect is seizable at small di-lepton mass and can enhance the $c\to ul^+l^-$ branching ratio by up to a factor of three. 

 \section{Long distance contributions}

The most serious problem with looking for new physics in charm processes is the presence of the long distance (LD) contributions in the corresponding meson decays. 

\begin{figure}[h]
\input{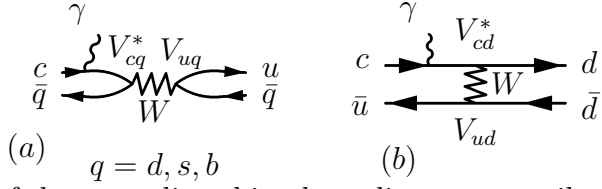}
\caption{The sketch of the most disturbing long distance contributions to the meson decays with  flavour structure (a) $c\bar q\to u\bar q\gamma$, $q=d,s,b$ and (b) $c\bar u\to u\bar u\gamma$, which are of interest to probe $c\to u\gamma$. The photon can be emitted from any quark line.\label{fig.ld}}   
\end{figure}

The charm meson decays $D\to V\gamma$ ($V=\rho,K^*,\omega,\phi$) with the flavour structure $c\bar q\to u\bar q\gamma$ ($q=d,s$) are   dominated by the LD mechanism sketched in  Fig. \ref{fig.ld}a \cite{radiative.exp,radiative}. This disturbing LD contribution is much smaller in the case of the $b$ spectator quark due to the small $V_{cb}^*V_{ub}$. 
As $B_c\to B_u^*\gamma$ is the least LD contaminated decay, it has been proposed as a probe for the $c\to u\gamma$ transition  \cite{bc}. It's branching ratio can reach up to $1\times 10^{-6}$  is general MSSM \cite{sasa}, while the 
 standard model prediction \cite{bc}  is $1\times 10^{-8}$.

The long distance mechanism  $c\bar u\to d\bar d\gamma$ (see Fig. \ref{fig.ld}b) overshadows the short distance 
process of interest $c\bar u\to u\bar u\gamma$ in 
$D^0\to \rho^0\gamma$ and $D^0\to \omega \gamma$ decays, 
since the final vector mesons $\rho^0$ and $\omega$ are 
mixtures of $u\bar u$ and $d\bar d$ states. The LD 
contributions largely cancel in the difference of the decays 
rates
\begin{equation} 
R=\frac{Br[D^0\to \omega\gamma]-Br[D^0\to \rho^0\gamma]}{Br[D^0\to \omega\gamma]}\propto Re\frac{{\cal A}[D^0\to u\bar u\gamma]}{{\cal A}[D^0\to d\bar d\gamma]}~,
\end{equation}
which is proportional to the short distance amplitude ${\cal A}[D^0\to u\bar u\gamma]$ driven by $c\to u\gamma$. 
The difference of the decay rates was proposed as a  probe for the $c\to u\gamma$ transition \cite{FPSW}. The relative difference $R$ can amount up to ${\cal O}(1)$ in general MSSM, as compared to the standard model prediction of $R=6\pm 15\%$ \cite{FPSW}.

\vspace{0.1cm}

\begin{table}[!htb]
\begin{center}
\begin{tabular}{|c|c c|}
\hline
 & $Br^{theor}$  &    $Br^{exp}$\\
\hline 
$ D^0 \to {\bar K}^{0} \mu^+\mu^-$ &$5.0\cdot 10^{-7}$&$<2.6\cdot 10^{-4}$\\
 $ D_s^+ \to \pi^+ l^+l^-$&$6.2\cdot 10^{-6}$&$<1.4\cdot 10^{-4}$\\
$ D^0 \to \pi^{0}l^+l^-$ &$3.9\cdot 10^{-7}$&$<1.8\cdot 10^{-4}$\\
$ D^0 \to \eta l^+l^-$ &$1.4\cdot 10^{-7}$&$<5.3\cdot 10^{-4}$\\
$ D^+ \to \pi^+ l^+l^-$&$1.7\cdot 10^{-6}$&$<7.8\cdot 10^{-6}$\\
$ D_s^+ \to K^{+ } l^+l^-$ &$8.3\cdot 10^{-8}$&$<1.4\cdot 10^{-4}$\\
$ D^+ \to K^{+} l^+l^-$&$8.2\cdot 10^{-9}$&$<8.1\cdot 10^{-6}$\\
$ D^0 \to K^{0} l^+l^-$&$1.3\cdot 10^{-9}$&\\
\hline
\end{tabular}
\caption{The predicted branching ratios $Br^{theor}$ \cite{pll} and the experimental upper bounds $Br^{exp}$ \cite{PDG,pll.exp} for the weak charm meson decays $D\to P \mu^+\mu^-$, which are  driven by the long distance mechanism.}
\label{tab.pll}
\end{center}
\end{table}

\begin{figure}[!htb]
\begin{center}
\includegraphics[scale=.4]{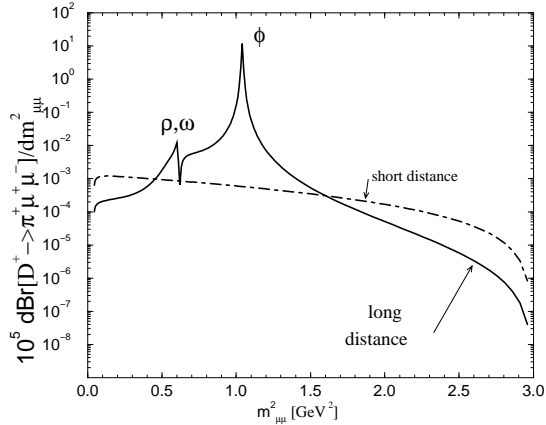} 
\caption{The differential branching ratio $dBr/dm_{\mu\mu}^2$ for $D^+\to \pi^+\mu^+\mu^-$ decay in the standard model. The dot-dashed line denotes the short-distance contribution driven by $c\to ul^+l^-$, while the solid line denotes the long distance contribution.  \label{fig.pll}} 
\end{center}  
\end{figure}

Similarly, the $c\to ul^+l^-$ transition is overshadowed by the LD contributions in the charm meson decays $D\to Vl^+l^-$ \cite{vll.exp,vll} and $D\to Pl^+l^-$ ($P=\pi,K,\eta$) \cite{pll.exp,pll}. The first theoretical analysis of all $D\to Pl^+l^-$ decays has undertaken the Heavy Meson Chiral Lagrangian approach \cite{pll} with the predictions given in Table \ref{tab.pll}.  
The unique possibility to look for $c\to ul^+l^-$ transition is represented by the $D\to \pi l^+l^-$ decays in the kinematical region of $m_{ll}$  above the resonance $\phi$, where the long distance contribution is reduced (see Fig. \ref{fig.pll}). The kinematical region above $\phi$ does not exist in other $D\to Xl^+l^-$ decays, since all the other mesons are heavier than the pion.  Unfortunately, this kinematical region is  not suitable to probe the possible enhancement of $c\to ul^+l^-$ within general MSSM, which enhances the rate at small $m_{ll}$ as displayed in Fig. \ref{fig.sd}. In general, the $D\to Pl^+l^-$ decays are  not suitable to probe the possible enhancement of $c\to u\gamma\to u l^+l^-$ decay via an almost real photon, since the decay $D\to P\gamma$ is forbidden for the case of the real photon.

 \section{Conclusions}

We  have examined the $c\to u\gamma$ and $c\to ul^+l^-$ rates in the context of the minimal supersymmetric standard model and proposed the relevant hadronic observables, which are relatively free of the long distance contributions. 

Schemes with universal soft breaking terms have negligible effect on these processes.  If the universality condition is relaxed, the dominant contribution comes from the gluino exchange diagrams with the left-right squark mass insertion. The $c\to u\gamma$ rate can be enhanced\cite{sasa} up to $6\times 10^{-6}$, compared to the QCD corrected standard model prediction  $3\times 10^{-8}$. The supersymmetric effect is seizable at small di-lepton masses in the case of the $c\to ul^+l^-$ decay and can enhance its rate by at most a factor of three. 

The supersymmetric  enhancement of the $c\to u\gamma$ rate can be  probed in the $B_c\to B_u^*\gamma$ decay \cite{bc} at LHC, which is expected to produce $2\times 10^{8}$ $B_c$ mesons with $p_T>20$ GeV. Another possibility is represented by  the difference of the decays rates $D^0\to \rho^0\gamma$ and $D^0\to \omega \gamma$ \cite{FPSW}, which could be explored at BaBar, Belle, BTeV and possibly tau-charm factory. The $c\to ul^+l^-$ transition is overshadowed by the long distance contributions to rare di-lepton charmed meson decays and we present the predictions for their rates in Table \ref{tab.pll}.  
 
\section*{Acknowledgments}

I would like to thank the organizers of the CICHEP  conference in Cairo for organizing this interesting meeting and the University of Trieste for the financial support.

\end{document}